\documentclass
[aps,prl,showpacs,onecolumn,superscriptaddress,amsfont,graphicx,nofootinbib]{revtex4}%
\usepackage{color,graphicx,epsfig}
\usepackage{ifpdf}
\usepackage{amsmath}
\usepackage{bm}
\usepackage{color}
\usepackage[english]{babel}
\usepackage{graphicx}%
\usepackage{amsfonts}%
\usepackage{amssymb}

\definecolor{Red}{rgb}{1.,0.,0.}

\begin{document}
\begin{flushright}%
TTP09-27\\SFB/CPP-09-71\\
\end{flushright}

\title{Tree-level contributions to the rare decays\\$B^{+}\rightarrow \pi^+\nu\bar{\nu}$, $B^{+}\rightarrow K^+\nu\bar{\nu}$, and $B^{+}\rightarrow K^{\ast+}\nu\bar{\nu}$ in the Standard Model}

\author{Jernej F. Kamenik} 
\email[Electronic address:]{jernej.kamenik@lnf.infn.it} 
\affiliation{INFN, Laboratori Nazionali di Frascati, Via E. Fermi 40, I-00044 Frascati, Italy}
\affiliation{J. Stefan Institute, Jamova 39, P. O. Box 3000, 1001
  Ljubljana, Slovenia}

\author{ Christopher Smith} 
\email[Electronic address:]{chsmith@particle.uni-karlsruhe.de} 
\affiliation{Institut f\"{u}r Theoretische Teilchenphysik, 
Karlsruhe Institute of Technology, D-76128 Karlsruhe, Germany}

\date{August 8, 2009}

\begin{abstract}
The tree-level contributions to the rare decays $B^{+}%
\rightarrow\pi^{+}\nu\bar{\nu}$, $B^{+}\rightarrow K^{+}\nu\bar{\nu}$, and $B^{+}\rightarrow K^{\ast+}\nu\bar{\nu}$ are analyzed and compared to those occurring in $K^{+}\rightarrow\pi^{+}\nu\bar{\nu}$, $D^{+}\rightarrow\pi^{+}\nu\bar{\nu}$, and $D_{s}^{+}\rightarrow\pi^{+}\nu\bar{\nu}$. It is shown that these purely long-distance contributions, arising from the exchange of a charged lepton, can be significant in $B^+$ decays for an intermediate $\tau$, potentially blurring the distinction between the modes used to extract $B^{+}\rightarrow\tau^{+}\nu_{\tau}$ and those used to probe the genuine short-distance $b\rightarrow d\nu\bar{\nu}$ and $b\rightarrow s\nu\bar{\nu}$ FCNC transitions. Numerically, the tree-level contributions are found to account for 97\%, 12\% and 14\% of the total $B^{+}\rightarrow\pi^{+}\nu\bar{\nu}$, $B^{+}\rightarrow K^{+}\nu\bar{\nu}$, and $B^{+}\rightarrow K^{\ast+}\nu\bar{\nu}$ rates, respectively.

\end{abstract}


\maketitle

\section{Motivation}

The rare semileptonic decays with neutrinos in the final state are considered as excellent probes of New Physics (NP) as they are significantly suppressed in the Standard Model (SM) and their long-distance (LD) contributions are generally subleading. By contrast, rare decays involving charged leptons in the final state often receive a substantial contribution from the photon penguin, which is not so effective at suppressing LD effects. Recently, the theoretical precision of the SM predictions for the $B\rightarrow K^{(\ast)}\nu\bar{\nu}$ modes has been improved to below $15\%$ \cite{Altmannshofer:2009ma} and already matches the expected sensitivity of a future super flavour factory \cite{Hitlin:2008gf,CKM2008}. Similarly, the theoretical precision achieved for the $K\rightarrow\pi\nu\bar{\nu}$ modes is now below 5\%, excluding parametric uncertainties from the CKM matrix elements, and a correspondingly precise measurement is expected to come out of the next generation of kaon experiments at CERN and J-PARC (see e.g. \cite{CKM2008}).

For all rare charged meson decays, there is a possible long-distance
contribution arising at tree-level through an intermediate lepton state. Such contributions are potentially large when the intermediate lepton is kinematically allowed to be on-shell, as can be seen in the case of $B^{+}\to K^{+} \nu\bar\nu$ by multiplying the measured branching ratios of $B\rightarrow\tau\nu$ and $\tau\rightarrow K\nu$ \cite{pdg}:
\begin{equation}
Br(B^{+}\rightarrow K^{+}\nu\bar{\nu})_{\mathrm{Tree}}\approx Br(B^{+}\rightarrow\tau^{+}\nu)\times Br(\tau^{+}\rightarrow K^{+}\bar{\nu})=(1.0\pm0.3)\times10^{-6}\,,
\end{equation}
to be compared with the most precise SM prediction for the short-distance, FCNC-induced rate $Br(B^{+}\rightarrow K^{+}\nu\bar{\nu})_{\mathrm{SD}}=(4.5\pm0.7)\times10^{-6}$ \cite{Altmannshofer:2009ma}. The purpose of this letter is to analyze in detail such contributions for the $B^+\rightarrow \pi^{+}\nu\bar{\nu}$ and $B^+\rightarrow K^{(\ast)+}\nu\bar{\nu}$ decays. First, however, we will review the situation for the $K^{+}\rightarrow\pi^{+}\nu\bar{\nu}$ decay where the intermediate lepton is never on-shell  \cite{TreeKaon,IsidoriMS05}, as well as for the LD dominated $D^{+}$ and $D_{s}^{+}$ decays, for which such tree-level effects were only briefly considered previously \cite{Burdman:2001tf}.

\section{Tree-level contributions to the rare decays $P^{+}\rightarrow P^{\prime+}\nu_{\ell}\bar{\nu}_{\ell}$}

The tree-level contribution for $P^{+}\rightarrow P^{\prime+}\nu_{\ell}\bar{\nu}_{\ell}$ proceeds through the leptonic decay $P^{+}\rightarrow\ell^{+}\nu_{\ell}$ followed by the transition $\ell^{+}\rightarrow P^{\prime+}\bar{\nu}_{\ell}$, as depicted in Fig. 1. A straightforward computation gives the differential decay rate for the decay chain $P^{+}\left(  p\right)  \rightarrow \nu_{\ell}\left(  p_{\nu}\right) \ell^{+}\left(  p_{\ell}\right) [ \rightarrow P^{\prime+}\left(  k\right)  \bar{\nu}_{\ell}\left(  p_{\bar{\nu}}\right) ]$ as
\begin{equation}
\frac{d\Gamma(P^{+}\rightarrow P^{\prime+}\nu_{\ell}\bar{\nu}_{\ell
})_{\mathrm{Tree}}}{dq^{2}dp_{\ell}^{2}}= -\frac{\left|  G_{F}^{2}V_{ij}V_{kl}^{\ast}f_{P}f_{P^{\prime}}\right|  ^{2}}{64\pi^{3}m_{P}^{3}}p_{\ell}^{4}\frac{m_{P}^{2}\left(  m_{P^{\prime}}^{2}-p_{\ell}^{2}\right)  +p_{\ell}^{2}\left(  p_{\ell}^{2}+q^{2}-m_{P^{\prime}}^{2}\right)  }{(m_{\ell}^{2}-p_{\ell}^{2})^{2}+m_{\ell}^{2}\Gamma_{\ell}^{2}}\,,
\end{equation}
where $p_{\ell}=k+p_{\bar{\nu}}$, $q=p_{\nu}+p_{\bar{\nu}}$, $V_{ij}$ are appropriate CKM elements (which we take from the SM fit of CKMFitter \cite{Tisserand:2009ja} in numerical applications), and the decay constants are defined from $\langle0|\bar{q}\gamma_{\mu}\gamma_{5}q|P\left( p\right) \rangle=if_{P}p_{\mu}$. The intermediate lepton width $\Gamma_{\ell}$ has to be accounted for to regulate the divergence when the lepton pole is inside the phase-space, and is introduced using the usual substitution $m_{\ell}^{2}\rightarrow m_{\ell}^{2}-im_{\ell}\Gamma_{\ell}$.

\begin{figure}[t]
\begin{center}
\includegraphics[width=8cm]{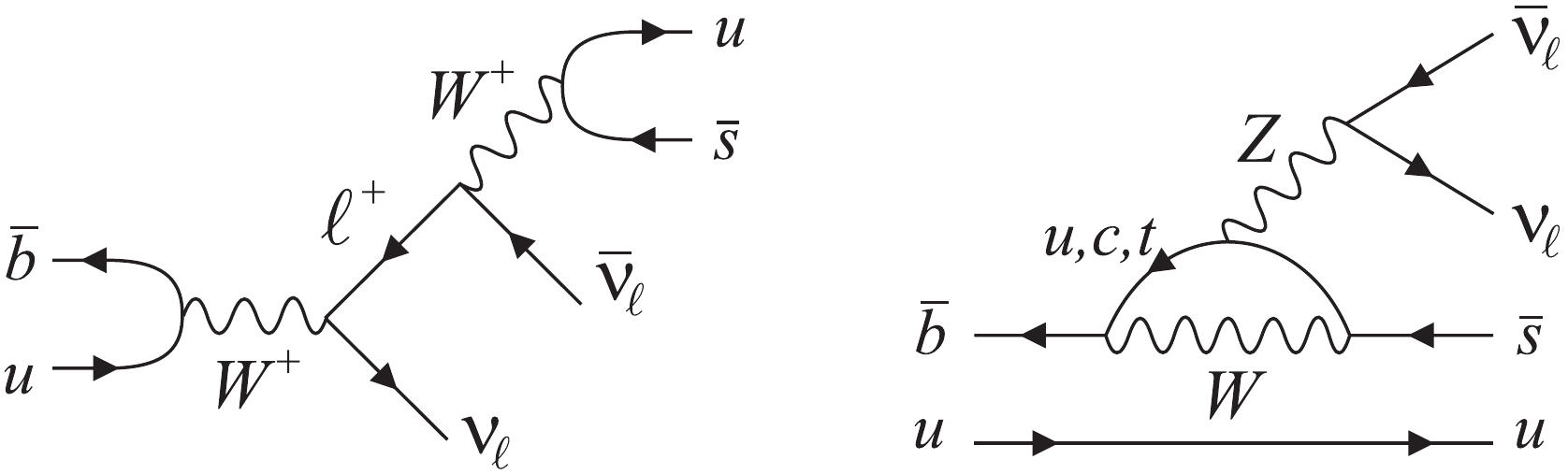}
\end{center}
\caption{ The tree-level charged-current process and the $Z$ penguin FCNC process (the $W$ box is understood) contributing to the rare charged meson decays, shown for $B^{+}\rightarrow K^{(\ast)+}\nu\bar{\nu}$ for definiteness.}%
\label{fig:2}%
\end{figure}

This contribution is formally of order $G_{F}^{4}$, i.e. of the same weak order as the loop-level FCNC contributions (see Fig. 1). However, the $Z$ penguin is dominated by the quadratic $SU(2)_{L}$ breaking, leading to an effective dimension-six operator, hence to an a priori larger contribution of $\mathcal{O}(G_{F}^{2}\alpha^{2})$ to the total rate. This na\"ive counting does not hold if the intermediate lepton can be on-shell, since the rate is then given to an excellent approximation by%
\begin{equation}
\Gamma(P^{+}\overset{}{\rightarrow}P^{\prime+}\nu_{\ell}\bar{\nu}_{\ell})_{\mathrm{Tree}}=\frac{\left|  G_{F}^{2}V_{ij}V_{kl}^{\ast}f_{P}f_{P^{\prime}}\right|  ^{2}}{256\pi^{3}m_{P}^{3}}\frac{2\pi m_{\ell}(m_{P^{\prime}}^{2}-m_{\ell}^{2})^{2}(m_{P}^{2}-m_{\ell}^{2})^{2}}{\Gamma_{\ell}}+\mathcal{O}(\Gamma_{\ell}^{0})\;. \label{Rate1}%
\end{equation}
With $\Gamma_{\ell}$ of order $G_{F}^{2}$, the tree-level contribution is of order $G_{F}^{2}$ and could become dominant.

The relative strength of the tree and loop contributions is very different in the case of the $K$, $D$ or $B$ meson decays, and we will now discuss them in turn.

\subsection{The rare decay $K^{+}\rightarrow\pi^{+}\nu\bar{\nu}$}

Since the $P^{+}\rightarrow\ell^{+}\nu_{\ell}$ process is helicity-suppressed, i.e. the amplitude is proportional to $m_{\ell}$, one could think that the $\tau$ lepton would give the largest contribution, the two $m_{\tau}$ factors from the vertices cancelling the $m_{\tau}^{2}$ of the $\tau$ propagator. However, for off-shell $\tau$, the helicity suppression is no longer effective: the $\tau$ momentum $p_{\tau}$ occurs instead of $m_{\tau}$, and since $p_{\tau}\sim\mathcal{O}(m_{K})\ll m_{\tau}$, the amplitude is suppressed by $\mathcal{O}(m_{K}^{2}/m_{\tau}^{2})$:%
\begin{equation}
\mathcal{M}\left(  K^{+}\left(  p\right)  \rightarrow\pi^{+}\left(  k\right)\nu_{\tau}\left(  p_{\nu}\right)  \bar{\nu}_{\tau}\left(  p_{\bar{\nu}}\right)  \right)  _{\mathrm{Tree}}=G_{F}^{2}V_{us}^{\ast}V_{ud}f_{K}f_{\pi}\frac{p_{\tau}^{2}}{p_{\tau}^{2}-m_{\tau}^{2}}\overline{u}_{\nu}\!\not \!k\left(  1-\gamma_{5}\right)  v_{\bar{\nu}}\;.
\end{equation}
This amplitude can be seen as deriving from an effective dimension-ten operator suppressed by $M_{W}^{4}m_{\tau}^{2}$. Numerically, this leads to a tiny $Br(K^{+}\rightarrow \pi^{+}\nu_{\tau}\bar{\nu}_{\tau})_{\mathrm{Tree}}\sim10^{-18}$ (using PDG values for the masses and decay constants \cite{pdg}), to be compared to the SD contribution from the $Z$ penguin and $W$ box of $\left(8.51\pm0.73\right)  \times10^{-11}$ in the SM \cite{IsidoriMS05,RareK}. The interference with the short-distance contribution is larger but still negligible, $Br(K^{+}\rightarrow\pi^{+}\nu\bar{\nu})_{\mathrm{Int.}}\sim10^{-15}$.

On the other hand, the contributions from the light leptons are not suppressed by a large mass scale. These effects where considered in Ref. \cite{IsidoriMS05}, along with chiral loop corrections, and amount to a small correction usually incorporated in $\delta P_{u,c}$ in the SM prediction for $K^{+}\rightarrow\pi^{+}\nu\bar{\nu}$. The tree-level exchanges are thus much smaller than the SD contributions. In fact, even the residual up-quark contribution to the $Z$ penguin gives a larger effect, see Ref. \cite{IsidoriMS05} for details.

\subsection{The rare decays $D_{(s)}^{+}\rightarrow\pi^{+}\nu\bar{\nu}$ and $D_{(s)}^{+}\rightarrow K^{+}\nu\bar{\nu}$}

The GIM suppression is very effective for $D^{+}\rightarrow \pi^{+}\nu\bar{\nu}$ and $D_{s}^{+}\rightarrow K^{+}\nu\bar{\nu}$, and makes their loop-level FCNC contributions extremely small. Further, the $Z$ penguin does not contribute to $D^{+}\rightarrow K^{+}\nu\bar{\nu}$ and $D_{s}^{+}\rightarrow\pi^{+}\nu\bar{\nu}$. Even including the LD contributions from vector mesons, the branching ratios for all these modes are tiny, typically below the $10^{-14}$ level \cite{Burdman:2001tf}. On the other hand, compared to $K\rightarrow \pi\nu\bar{\nu}$, the $\tau$ can now be on-shell and gives a large tree-level contribution. In fact, all the other contributions are so suppressed that $D^{+}\rightarrow\pi^{+}\nu\bar{\nu}$ and $D_{s}^{+}\rightarrow\pi^{+}\nu\bar{\nu}$ are used to measure the corresponding leptonic decays $D^{+}\rightarrow\tau^{+}\nu_{\tau}$ and $D_{s}^{+}\rightarrow\tau^{+}\nu_{\tau}$, since Eq. (\ref{Rate1}) can be written as
\begin{equation}
\Gamma(D_{(s)}^{+}\rightarrow\pi^{+}\nu_{\tau}\bar{\nu}_{\tau})_{\mathrm{Tree}}=\frac{1}{\Gamma_{\tau}}\Gamma(D_{(s)}^{+}\rightarrow\tau^{+}\nu_{\tau})\Gamma\left(  \tau^{+}\rightarrow\pi^{+}\bar{\nu}_{\tau}\right)  +\mathcal{O}(\Gamma_{\tau}^{0})\;.
\end{equation}
The kaon modes are not used for such measurements since $\tau^{+}\rightarrow K^{+}\bar{\nu}_{\tau}$ is Cabibbo-suppressed, but could in principle be included as well since there are no other significant contributions to worry about. It should be mentioned also that not only the hadronic $\tau^{+}\rightarrow\pi^{+}\bar{\nu}_{\tau}$ decay mode is looked for, but also purely leptonic decay channels.

\begin{figure}[t]
\begin{center}
\includegraphics[width=4cm]{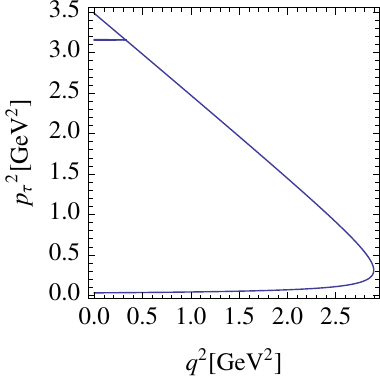}\;\;
\includegraphics[width=4cm]{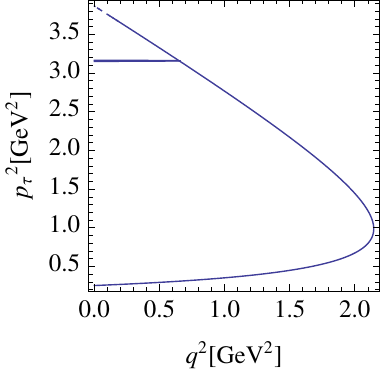}\;\;
\includegraphics[width=4cm]{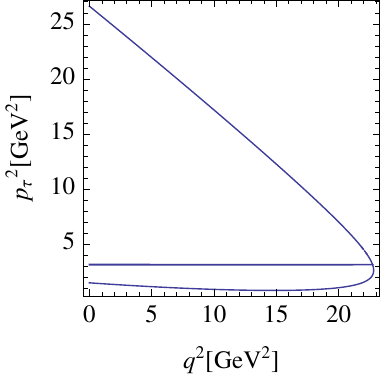}\;\;
\includegraphics[width=4cm]{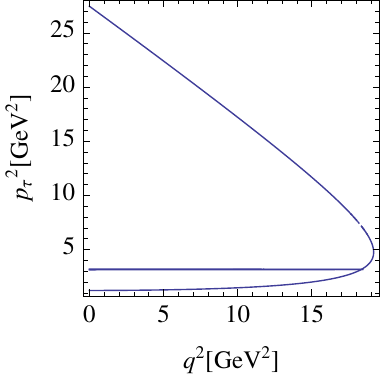}
\end{center}
\caption{From left to right: Contours of the Dalitz plots of the $D^{+}\rightarrow\pi^{+}\nu\bar{\nu}$, $D_{s}^{+}\rightarrow K^{+}\nu\bar{\nu}$, $B^{+}\rightarrow K^{+}\nu\bar{\nu}$, and $B^{+}\rightarrow K^{\ast+}\nu\bar{\nu}$ modes, with the horizontal lines denoting the position of the $\tau$ pole contribution. }%
\label{fig:1}%
\end{figure}

Let us stress that experimentally, the $\tau$ lifetime is not sufficient to filter out the tree-level transitions through the $\tau$ lepton pole. Further, the kinematics of the final state neutrinos is not measured, and the Dalitz plot of the $P^{+}\rightarrow P^{\prime+}\nu_{\tau}\bar{\nu}_{\tau}$ three-body decays cannot be reconstructed. A single kinematical variable, e.g. the invariant mass of the invisible particles $q^2$ alone (which is related to the $P^{\prime+}$ momentum), cannot always be used cut away the $\tau$ pole contributions since the $\tau$ exchange proceeds in the annihilation channel, which needs to be integrated over. For $D$ decays, however, there is not much phase-space available for creating an on-shell $\tau$ (see Fig. 2), and thus one could in principle disentangle the non-resonant contributions by restricting $q^2$ to large values (or equivalently, discarding events with a large $\pi^{+}$ or $K^{+}$ momentum). For instance, if it is restricted to $q^{2}>q_{cut}^{2}$ with%
\begin{equation}
q_{cut}^{2}=(m_{\tau}^{2}-m_{\pi}^{2})(m_{D}^{2}-m_{\tau}^{2})/m_{\tau}^{2}\;,\label{qcut}%
\end{equation}
the $\tau$ pole contribution is completely discarded.

Numerically, using the lattice estimates $f_{D^+}=(207\pm4)$ MeV and $f_{D_{s}^{+}}=(241\pm3)$ MeV \cite{Follana:2007uv}, we find%
\[%
\begin{array}
[c]{cccc}
& \ell=\tau & \ell=e,\mu\; & q_{cut}^2\\
Br(D^{+}\rightarrow\pi^{+}\nu_{\ell}\bar{\nu}_{\ell})\;\;\; & \;\;1.12(4)\times 10^{-4}\;\; & \;\;5\times10^{-16}\;\; & \;\;(0.58\;GeV)^2\;\;,\\
Br(D^{+}\rightarrow K^{+}\nu_{\ell}\bar{\nu}_{\ell})\;\;\; & \;\;7.2(3)\times 10^{-6}\;\; & \;\;2\times10^{-17}\;\; & \;\;(0.56\;GeV)^2\;\;,\\
Br(D_{s}^{+}\rightarrow\pi^{+}\nu_{\ell}\bar{\nu}_{\ell})\;\;\; &
\;\;5.3(1)\times10^{-3}\;\; & \;\;8\times10^{-15}\;\; & \;\;(0.84\;GeV)^2\;\;,\\
Br(D_{s}^{+}\rightarrow K^{+}\nu_{\ell}\bar{\nu}_{\ell})\;\;\; &
\;\;3.4(1)\times10^{-4}\;\; & \;\;4\times10^{-16}\;\; & \;\;(0.81\;GeV)^2\;\;.
\end{array}
\]
When a cut is enforced, the non-resonant $\tau$ contribution is of a similar size as those of the muon and electron, but shows a strong sensitivity to the precise value of $q_{cut}^{2}$. In any case, the experimental prospect for measuring such small branching ratios is remote, unless substantially enhanced by some NP contributions, so let us turn to $B$ decays.

\subsection{The rare decays $B^{+}\rightarrow \pi^{+}\nu\bar{\nu}$, $B^{+}\rightarrow K^{+}\nu\bar{\nu}$, and
$B^{+}\rightarrow K^{\ast+}\nu\bar{\nu}$}

The situation for $B$ decays is intermediate between that for
$K\rightarrow\pi\nu\bar{\nu}$ (full SD dominance) and for $D\rightarrow\pi\nu\bar{\nu}$ (full $\tau$ pole dominance). Indeed, the lepton pole contributions are, using $f_{B^+} = (200\pm20)$ MeV \cite{Lubicz:2008am},
\[%
\begin{array}
[c]{cccll}
& \ell=\tau & \ell=e,\mu & \;\;\;\;\;\mathrm{SD} & \;\mathrm{Full\;SM\;prediction}\\
Br(B^{+}\rightarrow\pi^{+}\nu_{\ell}\bar{\nu}_{\ell})\;\;\; & \;\;9.4(2.1)\times 10^{-6}\;\; & \;\;3\times10^{-17}\;\; & \;\;\sim3\times10^{-7}\;\; & \;\;9.7(2.1)\times10^{-6}\;\;\;,\\
Br(B^{+}\rightarrow K^{+}\nu_{\ell}\bar{\nu}_{\ell})\;\;\; & \;\;6.1(1.3)\times 10^{-7}\;\; & \;\;2\times10^{-18}\;\; & \;\;4.5(0.7)\times10^{-6}\;\; & \;\;5.1(0.8)\times10^{-6}\;\;,\\
Br(B^{+}\rightarrow K^{\ast+}\nu_{\ell}\bar{\nu}_{\ell})\;\;\; &
\;\;1.2(3)\times10^{-6}\;\; & \;\;5\times10^{-18}\;\; & \;\;7.2(1.1)\times10^{-6}\;\; & \;\;8.4(1.4)\times10^{-6}\;\;.
\end{array}
\]
The SD contributions are taken from Ref. \cite{Altmannshofer:2009ma}, rescaled when needed to account for the $B^0$ and $B^+$ lifetime difference, except for the pion mode which is estimated simply by rescaling $Br(B^{+}\rightarrow K^{+}\nu\bar{\nu})$ by $|V_{td}/V_{ts}|^{2}$ and correcting for the larger phase-space. For $B^{+}\rightarrow K^{\ast+}\nu_{\ell}\bar{\nu}_{\ell}$, the differential rate from the $\tau$ pole is, using $\langle0|\bar{s}\gamma_{\mu}u|K^{\ast}\left(  p,\epsilon\right)  \rangle=if_{K^{\ast}}m_{K^{\ast}}\epsilon_{\mu}$ with the QCD sum rules value $f_{K^{\ast}}=(220\pm5)$ MeV \cite{Ball:2006eu},%
\begin{equation}
\frac{d\Gamma(B^{+}\rightarrow K^{\ast+}\nu_{\tau}\bar{\nu}_{\tau
})_{\mathrm{Tree}}}{dq^{2}dp_{\tau}^{2}}=-\frac{\left|  G_{F}^{2}V_{ub}V_{us}^{\ast}f_{B}f_{K^{\ast}}\right|  ^{2}}{64\pi^{3}m_{B}^{3}}p_{\tau}^{4}\frac{(m_{K^{\ast}}^{2}-p_{\tau}^{2})(m_{B}^{2}-p_{\tau}^{2})+q^{2}(p_{\tau}^{2}-2m_{K^{\ast}}^{2})}{(m_{\tau}^{2}-p_{\tau}^{2})^{2}+m_{\tau}^{2}\Gamma_{\tau}^{2}}\,.
\end{equation}
In principle, there could be a sizable interference between the SD and LD contributions. However, the $\tau$ resonance is extremely narrow and often completely contained inside the Dalitz plot. When integrating over the $p_{\tau}$ variable, the SD part is fairly flat, with no appreciable phase shifts compared to the LD part. Therefore the resonance phase shift around the $\tau$ pole integrates the interference contribution almost to zero (it is of the order of $10^{-11}$ for $B\rightarrow K\nu\bar{\nu}$).

Because the rare $B^{+}$ decay modes can be used either to measure $B^{+}\rightarrow \tau^{+}\nu_{\tau}$ or to probe FCNC transitions, one has to decide how to deal with them on a case-by-case basis. 

Experimentally, the mode $B^{+}\rightarrow\pi^{+}\nu_{\ell}\bar{\nu}_{\ell}$ has been observed and used to extract the $B^{+}\rightarrow\tau^{+}\nu_{\tau}$ rate. This appears to be safe since compared to $B^{+}\rightarrow K^{+}\nu_{\ell}\bar{\nu}_{\ell}$, the SD amplitude is Cabibbo-suppressed while the tree-level amplitude is Cabibbo-favoured, resulting in a relative enhancement of LD with respect to SD by a factor $\sin^{-4}\theta_{c}\approx400$. However, note that the $\tau$ pole contribution is only about $97\%$ of the total $B^{+}\rightarrow\pi^{+}\nu_{\ell}\bar{\nu}_{\ell}$ rate in the SM. If the SD piece were enhanced by NP contributions, it would show up as a discrepancy between the $Br(B^{+}\rightarrow\tau^{+}\nu_{\tau})$ measured using $\tau^{+}\rightarrow\pi^{+}\bar{\nu}_{\tau}$ and other $\tau$ decay channels like $\tau^{+}\rightarrow e^{+}\nu_{e}\bar{\nu}_{\tau}$ or $\tau^{+}\rightarrow\mu^{+}\nu_{\mu}\bar{\nu}_{\tau}$ where there is no issue of entanglement with a SD contribution (still, the number of final state neutrinos is not measured so processes with identical charged leptons and hadrons but different numbers of neutrinos may be difficult to disentangle experimentally).

On the other hand, the $B^{+}\rightarrow K^{(\ast)+}\nu_{\ell}\bar{\nu}_{\ell}$ modes should not be used to measure the $B^{+}\rightarrow\tau^{+}\nu_{\tau}$ rate. In fact, one would rather want to remove the $\tau$ contribution as it is obscuring the interesting short-distance physics, and potential signals of NP. This is however difficult. Compared to the $D$ decays discussed in the previous section, there is no way to cut away the $\tau$ pole contribution using the invisible invariant mass $q^2$ in $B\rightarrow K^{(\ast)}\nu\bar{\nu}$ decays, as can be seen from Fig. 2. The best one can do is to cut away the low $q^2$ region (or high $K^{(\ast)}$ momentum) where the $\tau$ pole effect is the strongest, but a sizeable residual $\tau$ contribution is unavoidable.

\begin{figure}[t]
\begin{center}
\includegraphics[width=9cm]{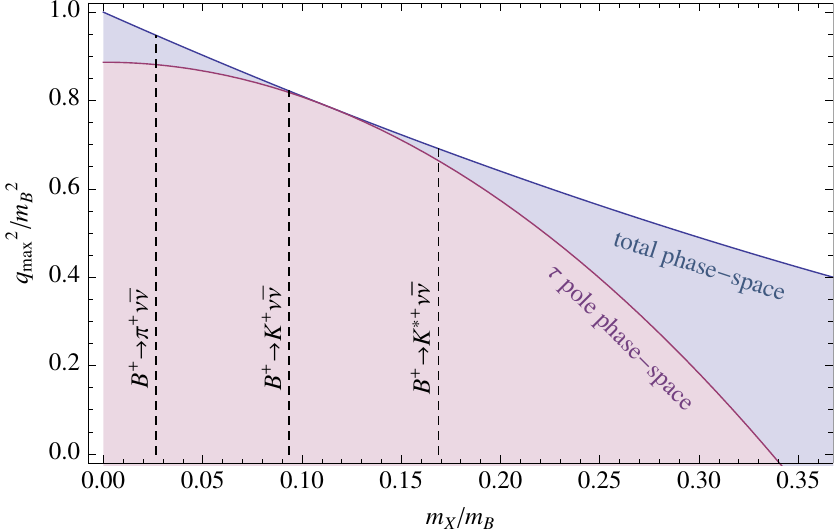}
\end{center}
\caption{Comparison of the total available phase-space in $B\rightarrow X\nu\bar{\nu}$ (denoted simply by the $\nu\bar{\nu}$ invariant mass $q^{2}$) with that where the $\tau$ can be on-shell, as a function of the total invariant mass $m_X$ of the visible decay products $X$.}%
\label{fig:3}%
\end{figure}

The kinematical configurations of the $B^{+}\rightarrow K^{(\ast)+}\nu_{\ell}\bar{\nu}_{\ell}$ decays are actually the worst possible to disentangle the SD and LD contributions. In Fig. 3 is shown the maximal kinematically allowed $q^2$ together with $q_{cut}^{2}$ of Eq. (\ref{qcut}) for a generic $B^{+}\rightarrow X\nu_{\ell}\bar{\nu}_{\ell}$ decay, as a function of the invariant mass of the $X$ state. It is only when this invariant mass is sufficiently large that the $\tau$ pole contribution can be cut away while still leaving a significant portion of phase-space to probe the SD contribution. In the extreme situation where the $X$ invariant mass is larger than $m_{\tau}$, the $\tau$ can never be on-shell and its contribution is negligible. Of course, for such a large invariant mass, experimentally reconstructing the decay is probably too difficult, while the SD contribution is significantly suppressed by the smaller matrix elements for $B\rightarrow X$. Therefore, the feasibility of this strategy remains to be seen, and for the time being, the $\tau$ pole contribution has to be considered as an irreducible background when probing the FCNC transition $b\rightarrow s\nu\bar{\nu}$ with charged $B$ decays.\footnote{Alternatively, one could probe the $b \rightarrow s \nu \bar{\nu}$ transition with $B_c^+ \rightarrow D_s^+ \nu \bar{\nu}$ for which the $\tau$ can never be on-shell. With a branching ratio around $10^{-6}$ \cite{bc}, the non-resonant $\tau$ contribution can be safely neglected.}

Finally, it should be mentioned that the $\tau$ pole contribution suffers from significant parametric uncertainties due to our poor knowledge of $V_{ub}$ and $f_{B}$. Fortunately, this uncertainty can be reduced in the SM by normalizing the effect to the neutral $B$ meson mass difference $\Delta m_d$ and using the corresponding hadronic bag parameter from the lattice instead of the decay constant \cite{Buras:2003td}. The relation is however not universal in presence of NP, therefore we do not apply it here. Instead, in view of future improvements in the determination of $|V_{ub}|$ and $f_{B}$, in particular from the $B\rightarrow\pi\nu\bar{\nu} $ mode, we give parametric formulas for the LD contributions to $B^{+}\rightarrow K^{(\ast)+}\nu\bar{\nu}$:
\begin{subequations}
\begin{align}
Br(B^{+}\rightarrow K^{+}\nu\bar{\nu})_{LD}  &  =0.606\times10^{-6}\left(\frac{f_{B}}{0.200\mathrm{GeV}}\right)^{2}\left(  \frac{f_{K}}{0.1555\mathrm{GeV}}\right)^{2}\left(  \frac{|V_{ub}|}{3.5\times10^{-3}}\right)^{2}\left(  \frac{|V_{us}|}{0.22521}\right)^{2}\,,\\
Br(B^{+}\rightarrow K^{\ast+}\nu\bar{\nu})_{LD}  &  =1.20\times10^{-6}\left(\frac{f_{B}}{0.200\mathrm{GeV}}\right)^{2}\left(  \frac{f_{K^{\ast}}}{0.220\mathrm{GeV}}\right)^{2}\left(  \frac{|V_{ub}|}{3.5\times10^{-3}}\right)^{2}\left( \frac{|V_{us}|}{0.22521}\right)  ^{2}\,.
\end{align}

\section{Conclusion}

In this letter, we have performed a systematic study of the tree-level processes contributing to the rare exclusive semi-leptonic decays of the charged $K$, $D$, and $B$ mesons into a neutrino pair. Our main result is that the contributions of the decay chains $B^{+}\rightarrow\nu_{\tau}\tau^{+}[\rightarrow K^{(\ast)+}\bar{\nu}_{\tau}]$ are large, about 15\% of the corresponding SM short-distance FCNC contributions from which they cannot be disentangled. These effects must be included in the SM predictions, which we have updated accordingly. On the other hand, $Br(B^{+}\rightarrow\pi^{+}\nu\bar{\nu})$ is dominated by the $\tau$ lepton pole, with the SD contribution being of about 3\%. As this mode is used to extract the $B^{+}\rightarrow\tau^{+}\nu_{\tau}$ rate, this pollution by SD physics should be kept in mind in case new physics turns out to be significant in $b\rightarrow d\nu\bar{\nu}$ transitions.

\subsection*{Acknowledgments}

This study was initiated during the Flavianet workshop \textit{Low energy constraints on extensions of the Standard Model} held on 23-27 July 2009 in Kazimierz, Poland, and the authors would like the thank the organizers for their hospitality. J. F. K. acknowledges useful discussions with A. Buras, T. Feldmann, and O. Cata. This work is supported by the European Commission RTN network, Contract No.
MRTN-CT-2006-035482 \textit{FLAVIAnet}, and by project C6 of the DFG Research Unit SFT-TR9 \textit{Computergest\"{u}tzte Theoretische Teilchenphysik}.

\end{subequations}

\end{document}